\documentclass[letterpaper]{article} 
\usepackage{aaai25}  
\usepackage{times}  
\usepackage{helvet}  
\usepackage{courier}  
\usepackage[hyphens]{url}  
\usepackage{graphicx} 
\usepackage{natbib}  
\usepackage{caption} 
\usepackage{xcolor}
\usepackage{booktabs}
\usepackage{multirow}
\usepackage{amsmath}

\usepackage{algorithm}
\usepackage{algorithmic}

\usepackage{newfloat}
\usepackage{listings}
\DeclareCaptionStyle{ruled}{labelfont=normalfont,labelsep=colon,strut=off} 
\floatstyle{ruled}
\newfloat{listing}{tb}{lst}{}
\floatname{listing}{Listing}
\usepackage{tikz}
\usetikzlibrary{positioning, arrows.meta}

\copyrighttext{}\title{Who Will Top the Charts? Multimodal Music Popularity Prediction via Adaptive Fusion of Modality Experts and Temporal Engagement Modeling}

\author{
    Yash Choudhary\textsuperscript{\rm 1},
    Preeti Rao\textsuperscript{\rm 2},
    Pushpak Bhattacharyya\textsuperscript{\rm 3}
}
\affiliations{
    \textsuperscript{\rm 1}Indian Institute of Technology Bombay, Mumbai, India\\
    \textsuperscript{\rm 2}Department of Electrical Engineering, IIT Bombay, India\\
    \textsuperscript{\rm 3}CFILT, Department of Computer Science, IIT Bombay, India\\
    200100173@iitb.ac.in, prao@ee.iitb.ac.in, pb@cse.iitb.ac.in
}

\begin{document}

\maketitle

\begin{abstract}
Predicting a song's commercial success prior to its release remains an open and critical research challenge for the music industry. Early prediction of music popularity informs strategic decisions, creative planning, and marketing. Existing methods suffer from four limitations: (i) temporal dynamics in audio and lyrics are averaged away; (ii) lyrics are represented as bag-of-words, disregarding compositional structure and affective semantics; (iii) artist- and song-level historical performance is ignored; and (iv) multimodal fusion approaches rely on simple feature concatenation, resulting in poorly aligned shared representations. To address these limitations, we introduce GAMENet, an end-to-end multimodal deep learning architecture for music popularity prediction. GAMENet integrates modality-specific experts for audio, lyrics, and social metadata through an adaptive gating mechanism. We use audio features from Music4AllOnion processed via OnionEnsembleAENet, a network of autoencoders designed for robust feature extraction; lyric embeddings derived through a large language model pipeline; and newly introduced Career Trajectory Dynamics (CTD) features that capture multi-year artist career momentum and song-level trajectory statistics. Using the Music4All dataset (113k tracks), previously explored in MIR tasks but not popularity prediction, GAMENet achieves a 12\% improvement in $R^2$ over direct multimodal feature concatenation. Spotify audio descriptors alone yield an $R^2$ of 0.13. Integrating aggregate CTD features increases this to 0.69, with an additional 7\% gain from temporal CTD features. We further validate robustness using the SpotGenTrack Popularity Dataset (100k tracks), achieving a 16\% improvement over the previous baseline. Extensive ablations confirm the model's effectiveness and the distinct contribution of each modality.

\end{abstract}



\section{Introduction}
The global recorded-music market generated \textbf{\$29.6\,billion in 2024}\footnote{\url{https://www.reuters.com}} while serving \textbf{752 million paying subscribers} who accessed over \textbf{202 million tracks} across streaming platforms.  Yet approximately \textbf{86.9\%} of those tracks failed to attain Spotify’s threshold of \textit{1000 annual plays}, and nearly \textbf{95\%} of artists attracted fewer than \textit{1000 monthly listeners}\footnote{ \url{https://chaoszine.net}}. This stark disparity motivates the importance of predicting song popularity \textit{prior to release} to enable efficient resource allocation, targeted promotion, and informed creative decisions in an increasingly competitive streaming landscape.

Music popularity prediction, formally known as Hit Song Science (HSS) since the early 2000s \cite{Seufitelli2023HitSS}, examines the factors influencing a song's commercial success using intrinsic attributes such as audio features, lyrical content, and artist profiles, alongside extrinsic signals including streaming trends, listener engagement, and cultural factors. Popularity modeling typically targets success metrics like chart appearances, peak rankings, and user engagement data such as likes, downloads, shares, and Spotify popularity scores \cite{Seufitelli2023HitSS}. Prior work has explored a variety of modeling strategies for music popularity prediction, ranging from unimodal approaches based on audio features to multimodal architectures that incorporate lyrics, metadata, and user interaction signals. A review of these methods is presented in Section~\ref{sec: related works}.

Despite advances in multimodal modeling and large-scale datasets, important gaps persist in the feature coverage. Existing methods typically overlook temporal dynamics, using static aggregate representations for audio and lyrics. Lyric representations often rely on simple bag-of-words models, neglecting compositional semantics. Additionally, historical artist and song performance data, critical for capturing career momentum, is often neglected. Further, multimodal fusion often relies on simple feature concatenation, leading to poorly aligned representations and imbalanced modality contributions. Finally, current deep learning models lack interpretability, limiting their actionable insights for creative and commercial decisions.

We focus on addressing a subset of these limitations through the following main contributions:
\begin{enumerate}
    \item We introduce Career Trajectory Dynamics (CTD), a novel feature set capturing artist career trends and song trajectories. Statistical CTD features alone significantly improve performance ($R^2$ from 0.13 to 0.69), and incorporating temporal CTD features further improves results by 7\%, demonstrating the benefit of temporal feature modeling.
    
    \item We propose GAMENet (Gated Adaptive Modality Experts Network), a multimodal deep learning architecture with modality-specific experts integrated via a gated adaptive fusion mechanism, enabling effective cross-modal learning and interpretability. GAMENet achieves a 12\% improvement in $R^2$ compared to direct feature concatenation methods on Music4All and outperforms the current baseline on the SpotGenTrack Popularity Dataset (100k+ tracks)  by 15\%.

    \item In the course of researching new features, we work with the Music4All family with it large multimodal set of song characteristics and present the first ever music popularity prediction study on this dataset --- comprising Music4All and Music4All-ONION --- with 113k tracks and approximately 252 million user listening events to predict music popularity scores in the range 0--100.
\end{enumerate}

The remainder of this paper is structured as follows: Section~\ref{sec: related works} surveys related work; Section~\ref{sec:dataset-family} describes the datasets; Section~\ref{sec: methodology} outlines the proposed methodology; Section~\ref{sec:exp} presents experimental results and analysis; and Section~\ref{sec: conclusion} concludes the paper.

\section{Related Work}\label{sec: related works}

Hit Song Science (HSS) uses ideas from music information retrieval (MIR), machine learning, and social signal analysis to predict whether a song will become popular. Early research focused on handcrafted audio features and traditional machine learning models 
\cite{Dhanaraj2005,Pachet2008}. The release of large-scale datasets—such as the Million Song Dataset \cite{Bertin2011}, Last.fm 360K \cite{Celma2010}, and Spotify charts—allowed researchers to include user listening data and social signals in their models. These datasets also made it possible to add more detailed audio features (like MFCCs and spectral statistics), which helped improve prediction models \cite{Araujo2019,Shulman2016}.

Later work added more data types to these models, such as song lyrics. 
Metadata—like genre, artist popularity, and chart history—has also been useful. Social signals, especially patterns in how users listen to music, have led to major gains. For example, finding users with similar music tastes can improve prediction accuracy by up to 50\% \cite{Reisz2024Quantifying}. Other studies have looked at artist collaborations and how connected an artist is in the network, which also helps explain success \cite{Silva2021Collaboration}.

Another line of research shows that looking at a song or artist’s early performance over time can help predict long-term success. Some studies found that the first few weeks or months after release can give strong clues about a track’s future \cite{Chon2006,Lee2015}. Combining data from YouTube—such as views, likes, and comments—with audio features has also improved results \cite{Yee22}. Some methods use tools like canonical correlation analysis to combine social and content data \cite{Matsumoto2020Context}, while others rely on historical streaming and chart data to model career momentum \cite{Araujo2017,Araujo2019}.

More recently, deep learning models have shown strong performance by combining data from different sources \cite{hitmusicnet}. For example, wide-and-deep networks \cite{Zangerle2019Hit} and Siamese CNNs with ranking loss \cite{Yu2017Hit} are designed to compare songs based on their success. These advances have been helped by new datasets with rich annotations, such as SpotGenTrack \cite{hitmusicnet}, HSP \cite{Vötter2022}, and Music4All \cite{Santana2020Music4AllAN}. Recent work also explores how to model changes over time in audio and user signals \cite{Vavaroutsos2024HSP,Li2021LSTMRPAAS}. 

While prior research has made strong progress in modeling music popularity using multimodal features \cite{Seufitelli2023HitSS}, key gaps remain. Most existing models either treat modalities in isolation or rely on simple fusion strategies that do not account for the differing importance of each modality across songs. In addition, although some recent work models temporal trends in engagement signals, there is little focus on systematically capturing long-term artist-level momentum and career progression—despite evidence that early performance strongly correlates with future success. This applies not only to artists but also to individual songs, where early patterns of listener engagement often signal long-run popularity outcomes. Our work addresses these open challenges by introducing structured Career Trajectory Dynamics (CTD) features and a gated fusion architecture (GAMENet) that adaptively combines modality-specific predictions.

\section{Dataset}\label{sec:dataset-family}

In this section, we introduce the Music4All dataset family, comprising the original Music4All (M4A)~\cite{Santana2020Music4AllAN} and its enriched extension, Music4All-Onion (M4A-O)~\cite{Music4All-Onion}. We then describe our data cleaning and preprocessing steps. Additionally, to evaluate the generalizability of the GAMENet architecture, we use the SpotGenTrack Popularity Dataset (SPD)~\cite{hitmusicnet} as detailed in \ref{subsec:spd data}

\subsection{Music4All Family}\label{subsec:music4all-family}

Music4All (M4A) is described in its original release as “a new music database which contains metadata, tags, genre information, 30-second audio clips, lyrics, and so on,” designed to provide content-rich benchmarks for MIR research~\cite{Santana2020Music4AllAN}. The corpus was assembled through three parallel scraping pipelines:

\begin{itemize}
\item \textbf{Audio.} Full tracks were sourced from YouTube, converted to lossless format, resampled at 44.1\,kHz, loudness-normalized, and center-cropped to a 30-second.
\item \textbf{Lyrics.} Song lyrics were obtained via the Musixmatch API, retaining only lines with language-detection confidence $\geq 0.90$, followed by unicode normalization.
\item \textbf{Metadata \& tags.} Sixteen high-level audio descriptors (e.g., \textit{danceability}, \textit{energy}, \textit{valence}) and Spotify popularity scores (range: [0,100]) were collected via the Spotify Web API. The popularity score corresponds to each song's value as of the end of 2020. Semantic tags and genre labels were sourced from Last.fm and Every Noise at Once.
\end{itemize}

After deduplication and pruning of incomplete entries per modality, the final dataset comprises 109,269 tracks by 16,269 artists spanning release years from 1920 to 2020. The popularity distribution, approximates a Gaussian curve with mean $\mu \approx 35$ and standard deviation $\sigma \approx 15$, indicating suitability as a balanced benchmark dataset. The median track release year is 2011, with a distribution skewed towards recent music. Music4All-Onion (M4A-O)~\cite{Music4All-Onion} extends M4A by enriching each track with additional content descriptors and collaborative signals. Features are grouped into five semantic layers: (1) Audio descriptors, including MFCCs, rhythm, tonality, Essentia features and openSMILE,  ComParE statistics; (2) Embedded metadata such as TF–IDF and word2vec lyric embeddings and emotion scores; (3) Expert-generated genre profiles; (4) User-generated Last.fm tag distributions; and (5) Visual embeddings from YouTube video frames (VGG-19, Inception-v3, ResNet-50). Across these five layers, M4A-O provides 26 distinct feature sets for the same 109,269 tracks in M4A. In addition, M4A-O provides a set of 252,984,396 listening records from 119,140 users, extracted from the online music platform Last.fm. Each record captures a user–track interaction event, timestamped to the second, and aligned to a subset of 56,512 tracks in the corpus. 

\begin{figure}[t]
\centering
\includegraphics[width=0.8
\columnwidth]{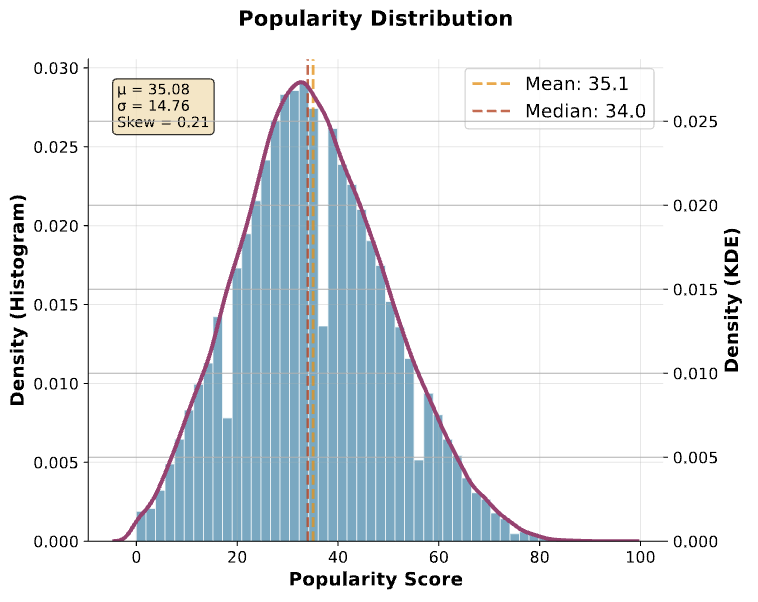}
\caption{Distribution of Spotify popularity scores in the original Music4All dataset.}
\label{fig:pop_dist_original}
\end{figure}

Together, M4A and M4A-O form a publicly licensed, large-scale multimodal dataset family that integrates raw and derived representations across audio and lyrics modalities, enriched with social metadata descriptors and large-scale user–track interaction logs.

\begin{figure}[t]
\centering
\includegraphics[width=0.8
\columnwidth]{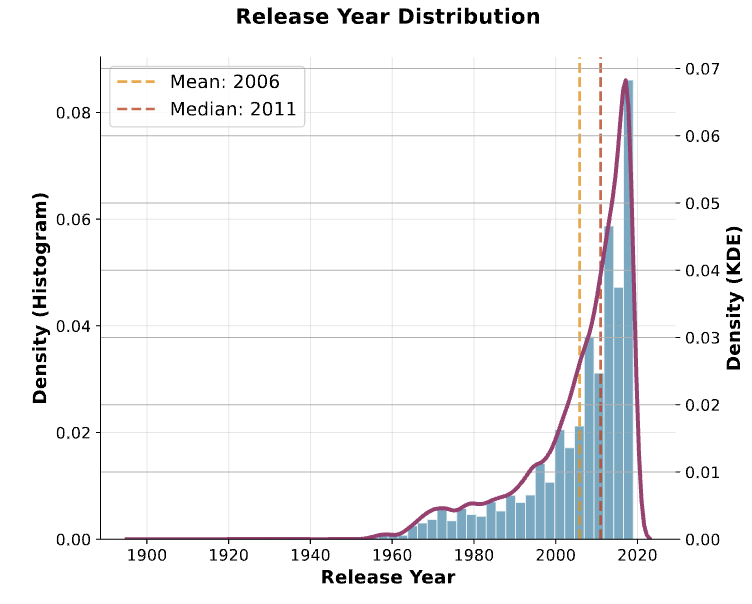}
\caption{Distribution of release years in the Music4All dataset, showing a sharp concentration in recent decades with mean 2006 and median 2011}
\label{fig:year_dist_original}
\end{figure}

\subsection{SpotGenTrack Popularity Dataset}\label{subsec:spd data}
The SpotGenTrack Popularity Dataset (SPD), introduced in~\cite{hitmusicnet}, comprises 101,939 tracks by 56,129 artists across 75,511 albums, collected via the Spotify and Genius APIs. The dataset aggregates top-50 playlists from 26 Spotify-available countries, capturing a geographically diverse sample of popular music. Each track is assigned a Spotify popularity score in the range [1,100], computed from platform-specific engagement metrics. These scores are approximately Gaussian distributed with mean $\mu = 40.02$ and standard deviation $\sigma = 16.79$, making the dataset well-suited for regression-based modeling. SPD includes multimodal features across three categories. High-level descriptors from Spotify (e.g., \textit{danceability}, \textit{valence}, \textit{tempo}) represent aggregate musical characteristics. Low-level audio representations such as MFCCs, chromagrams, and spectral features are extracted directly from the waveform. Lyrics are processed using stylometric analysis, yielding textual attributes like sentence complexity and vocabulary richness. Metadata includes artist popularity and geographic availability.
The multimodal structure of SPD supports both fine-grained content analysis and high-level trend modeling. Its combination of audio descriptors, stylometric lyric features, and artist-level metadata enables evaluation of models that integrate diverse input modalities. While not as extensive as M4A in interaction data, SPD provides a clean and well-curated benchmark for testing the generalizability of music popularity prediction architectures.

\subsection{Dataset Cleaning and Pre-processing}\label{subsec:data cleaning-preprocessing}
\textbf{SpotGenTrack Popularity Dataset} We filtered the SPD dataset to remove low-quality lyric entries, discarding tracks with lyrics shorter than 100 or longer than 7,000 characters, which typically contained noise such as placeholders or non-lyrical content. To ensure balanced representation, we retained only tracks in English, Spanish, Portuguese, French, or German, as other languages accounted for less than 1\% of the dataset. The final cleaned set consists of 51,319 English tracks and 22,887 tracks in the other selected languages. The distribution of popularity scores and release years remained consistent with the original dataset, ensuring no sampling bias was introduced.

\begin{figure}[t]
\centering
\resizebox{0.95\columnwidth}{!}{
\begin{tikzpicture}[
node distance=0.6cm and 1cm,
every node/.style={draw, rounded corners=2pt, align=center, font=\small, minimum width=2.8cm},
arrow/.style={draw, -{Stealth}, thick},
data/.style={fill=blue!10},
proc/.style={fill=orange!15},
filter/.style={fill=red!15},
final/.style={fill=green!20}
]

\node[data] (m4a) {Music4All\\109,269 tracks};
\node[filter, below=of m4a] (yearfilt) {Remove pre-1960 tracks\\(→ 108,854)};
\node[filter, below=of yearfilt] (langfilt) {Keep top 4 languages\\(→ 103,352)};
\node[proc, below=of langfilt] (norm) {Lyric normalization:\\ whitespace, tags, breaks};

\node[data, right=4.5cm of m4a] (onion) {Music4All-ONION\\253M listening events\\for 56,512 tracks (2010--2020)};
\node[proc, below=of onion] (groupby) {Group by Track ID\\and compute 4 per-year metrics};
\node[filter, below=of groupby] (dropnulls) {Drop feature columns\\for 2010--2015 due to sparsity};

\node[proc, below=-0.85cm of norm, xshift=5cm] (merge) {Merge filtered logs with\\ cleaned Music4All data};

\node[final, below=of merge] (final) {\textbf{M4A-CTD}\\53,777 tracks};

\draw[arrow] (m4a) -- (yearfilt);
\draw[arrow] (yearfilt) -- (langfilt);
\draw[arrow] (langfilt) -- (norm);

\draw[arrow] (onion) -- (groupby);
\draw[arrow] (groupby) -- (dropnulls);

\draw[arrow] (norm) -- (merge);
\draw[arrow] (dropnulls) -- (merge);
\draw[arrow] (merge) -- (final);

\end{tikzpicture}
}
\caption{Cleaning and processing pipeline from Music4All and Music4All-ONION to the final M4A-CTD dataset containing 53,777 tracks.}
\label{fig:cleaning-flow}
\end{figure}
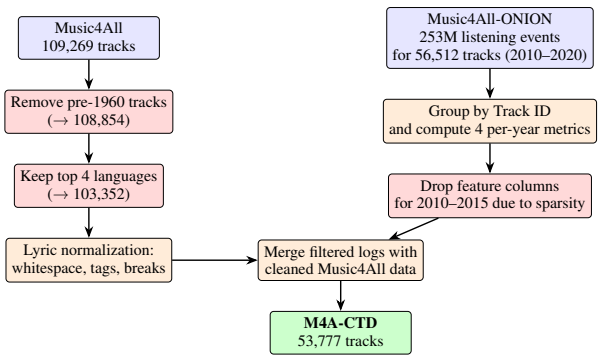

\textbf{Music4All Family} Prior to modality-specific feature engineering, we applied a structured cleaning pipeline to construct a high-quality subset of the Music4All corpus suitable for modeling. As shown in Figure~\ref{fig:cleaning-flow}, we began by filtering the Music4All dataset, which originally contained 109,269 tracks. A temporal filter was first applied, removing 478 tracks released before 1960 (0.44\%) due to their sparse presence. We then filtered by lyric language, retaining only the four most frequent categories: English (76.97\%), instrumental placeholders (8.62\%), Portuguese (6.42\%), and Spanish (2.95\%). Languages appearing in fewer than 1\% of tracks were excluded to mitigate class imbalance, resulting in a filtered set of 103,352 tracks (94.96\% of the post-1960 data). Each lyric entry was passed through a deterministic normalization pipeline to ensure consistent formatting and tokenization. Processing steps included trimming and collapsing whitespace, standardizing line breaks, expanding repetition markers (e.g., \texttt{[x2]}), and removing non-sung annotations such as \texttt{[Instrumental]}, \texttt{[Spoken]}, and \texttt{[Guitar Solo]}.
In parallel, we processed the Music4All-ONION dataset~\cite{Music4All-Onion}, which provides over 253 million timestamped listening events from 120,000 Last.fm users between 2010 and 2020, aligned with 56,512 unique track IDs in the Music4All data. These logs were grouped by track ID and aggregated into four per-year engagement metrics: total play count, unique play count, number of users with multiple plays (as a proxy for listener loyalty), and median play count per user (as a measure of engagement depth). Although the logs cover the full decade, the data from 2010 to 2015 had many missing values across most tracks. To address this, we dropped the feature columns corresponding to these early years and retained only the 2016--2020 feature segment for further processing. The aggregated listening logs were then merged with the filtered Music4All metadata using track IDs as keys. Only tracks with valid engagement data across the retained temporal window were included. This yielded the final cleaned dataset, \textbf{M4A-CTD}, comprising 53,777 tracks. Figure~\ref{fig:cleaning-flow} summarizes the full data cleaning pipeline. M4A-CTD closely resembles the original Music4All-Clean dataset in release year, popularity, and language distributions, indicating that key data properties were preserved during filtering.

To study the M4A-CTD dataset for music popularity prediction, we created a train-test split designed to ensure balanced representation across different popularity levels. Popularity scores were discretized into five quantile-based bins, and an 80/20 split was performed using stratification on these bins, with a fixed random seed (42) for reproducibility. This approach preserves the target distribution across splits and mitigates potential biases caused by class imbalance during model training and evaluation.

\section{Methodology}\label{sec: methodology}
\subsection{Career Trajectory Dynamics (CTD) Features}\label{subsec: Methodology CTD}

Listener engagement plays a key role in shaping music popularity. Previous studies have shown that listener interactions are important, but they often use static metrics without modeling temporal patterns~\cite{Shulman2016, Seufitelli2023HitSS}. To address this, we introduce a structured pipeline for extracting Career Trajectory Dynamics (CTD) features from aggregated listening logs in the M4A-CTD dataset, capturing engagement patterns at both song and artist levels.

We derive features to represent listener interactions at two levels. \textbf{Song-level features} include yearly metrics from 2016 to 2020: total plays, unique listeners, repeat listeners (users with multiple plays), and median plays per listener. Additionally, we compute song-level behavioral indicators such as the song loyalty rate (rate of repeat listening) and the song repeat ratio (frequency of repeated plays). \textbf{Artist-level features} capture career trends and engagement stability across an artist’s full track list. These include loyalty rate, loyalty growth rate, reach growth rate, loyalty consistency, and engagement consistency. We group these features into two categories: \textbf{CTD Aggregate Features:} Capture average and total engagement over the full five-year period at both song and artist levels. \textbf{CTD Temporal Features:} Preserve year-wise variation to capture momentum, growth, and engagement changes over time.
We evaluate the contribution of these features through ablation studies in Section~\ref{subsec: ctd ablation}.

\subsection{OnionEnsembleAENet}\label{subsec:OnionEnsembleAENet}
The Music4All-ONION dataset provides an extensive set of handcrafted audio features, spanning spectral patterns, emotional descriptors, pitch-based attributes, and statistical summaries. We select a refined subset of 11,851 features based on complementary informational content, excluding redundant categories such as overlapping MFCCs and repetitive voice statistics. To compress this heterogeneous high-dimensional feature space, we design OnionEnsembleAENet, a ensemble of seven autoencoders, each assigned to a compress semantically coherent group of features: Small Combined (439), BoW-Emobase-Chroma (1000), BLF Feature Group (4478), Essentia (1034), ComParE Audio Spectral (2800), ComParE MFCC (1400), and ComParE PCM (1700).
Each autoencoder follows a symmetric encoder-decoder architecture with a bottleneck layer to progressively reduce the dimensionality. The hidden layer are adaptively structured based on input dimensionality: inputs above 4000 dimensions use sequence [d/2, d/3, d/5]; inputs between 2000--4000 use [d/2, d/4]; and smaller groups use a single reduction layer [d/2].We apply ELU activations ($\alpha = 0.1$) in all layers except the bottleneck, alongside batch normalization and dropout ($p = 0.05$) to prevent overfitting. Hyperparameters including layer depth, activation functions, and regularization strategies were optimized to reduce reconstruction loss. The training objective for each autoencoder is a composite of reconstruction loss and latent space regularization, formalized as:
$$
\mathcal{L}_k = \text{MSE}(x_k, \hat{x}_k) + \lambda_k \| z_k \|_2^2
$$
where $x_k$ represents input features, $\hat{x}k$ reconstructed features, and $z_k$ the bottleneck embeddings. The regularization term $\lambda_k = 0.001 \times \frac{128}{d{\text{enc},k}}$ inversely scales with encoding dimensionality to control the latent representation norms. The compressed representations from all autoencoders are concatenated, forming a unified, low-dimensional audio embedding for integration into our multimodal prediction framework.

\begin{figure*}[t]
\centering
\includegraphics[width=1.8\columnwidth]{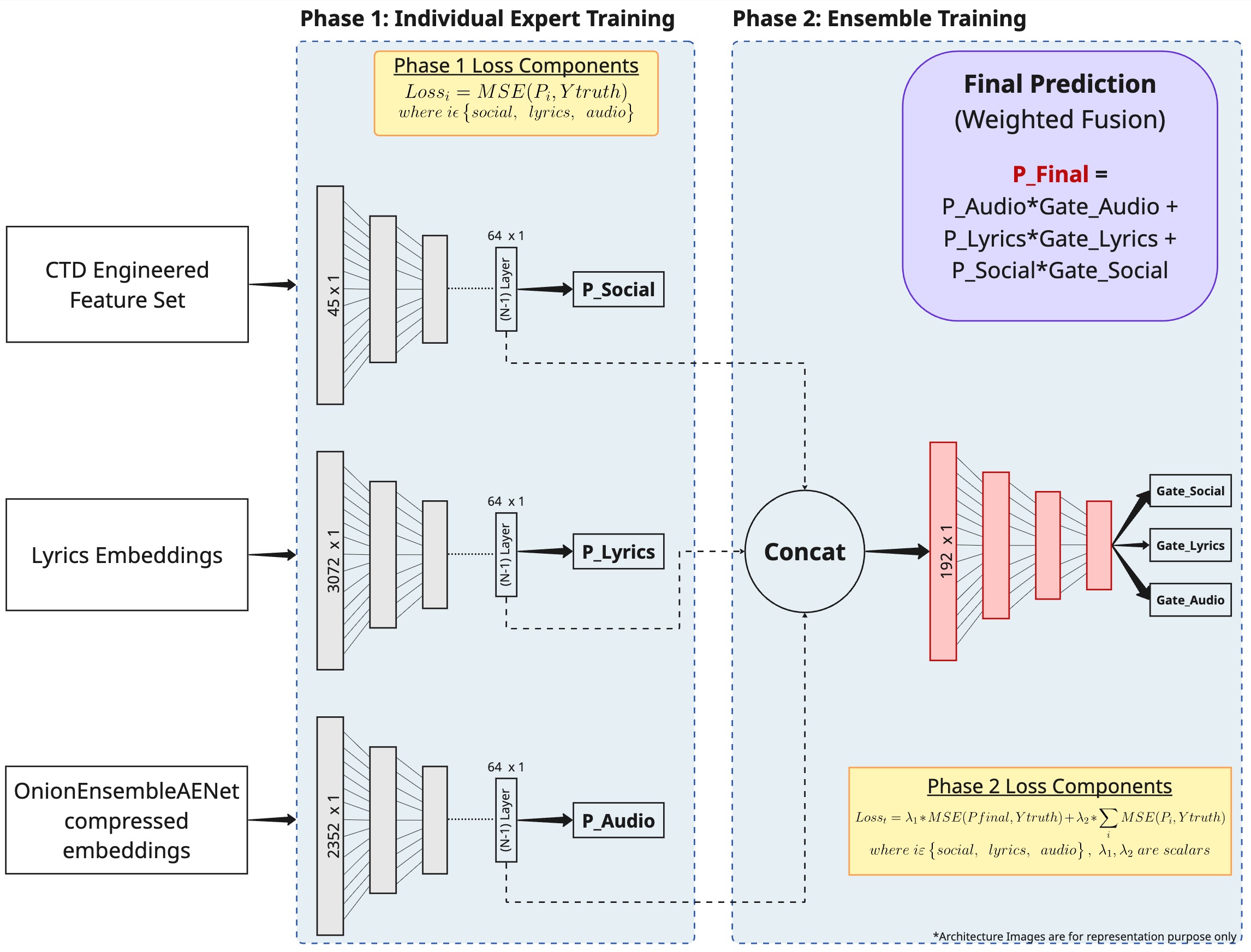}
\caption{GAMEnet}
\label{fig:gamenet}
\end{figure*}
\subsection{GAMENet: Gated Adaptive Modality Experts Network}\label{subsec:GAME-Net}
Integrating heterogeneous multimodal features for popularity prediction requires a principled approach that accounts for differences in feature scales, dimensions, and semantic content. To this end, we propose GAMENet ---  the Gated Adaptive Modality Experts Network  --- a multimodal ensemble model that integrates modality-specific deep networks with a learnable gating mechanism for adaptive fusion.

GAMENet operates on three distinct input modalities: compressed audio features from OnionEnsembleAENet (2,352 dimensions), OpenAI-generated text embeddings representing lyrics (3,072 dimensions), and structured social metadata (46 dimensions). Each modality is processed by a dedicated deep network branch, independently optimized to capture domain-specific patterns. The network architectures are empirically selected based on extensive hyperparameter tuning and cross-validation, balancing depth, activation choice, and regularization to maximize predictive performance.

The audio branch employs a four-layer feedforward network with hidden dimensions [512, 256, 128, 64], using ELU ($\alpha = 0.1$) activations, batch normalization, and progressively decreasing dropout rates [0.3, 0.2, 0.2, 0.1]. The lyrics branch, designed to handle high-dimensional semantic embeddings, uses a deeper network [1024, 512, 256, 128, 64] with similar regularization, ensuring sufficient capacity for capturing compositional semantics. The social metadata branch processes low-dimensional inputs through an expanded projection network [512, 256, 128, 64], employing LeakyReLU (slope = 0.05) activations with lighter regularization. This dimensional expansion balances the representation scale across modalities, preventing dominance effects during fusion. Each branch outputs a normalized popularity score in [0, 1] via a sigmoid-activated linear layer, consistent with the MinMax-scaled target values.

The core innovation of GAMENet is its learnable gating network, which adaptively combines modality-specific predictions based on intermediate feature representations. Each modality branch exposes its 64-dimensional penultimate layer output to the gating module. Before fusion, these features undergo learnable standardization to align magnitude scales:
$$
\tilde{x}_i = \frac{x_i - \mu_i}{|\sigma_i| + \epsilon}
$$
where $\mu_i$ and $\sigma_i$ are modality-specific, learnable parameters. The standardized features are concatenated and passed through a two-layer feedforward gating network [128, 64], with LeakyReLU activations, batch normalization, and light dropout (0.01). The final gating layer produces unnormalized attention logits, converted via softmax into attention weights $\alpha_i$ satisfying $\sum_i \alpha_i = 1$. The final ensemble prediction is computed as a weighted sum of individual modality outputs:
$$
\hat{y} = \sum_{i=1}^3 \alpha_i \cdot \sigma(f_i(x_i))
$$
This formulation enables GAMENet to adaptively weigh modalities per input sample, yielding interpretable and data-driven fusion behavior.

Training proceeds in two phases. Phase 1 optimizes each modality-specific branch independently using mean squared error (MSE) loss. Phase 2 jointly trains the gating network and optionally fine-tunes the modality branches using a composite loss:
$$
\mathcal{L}_{\text{total}} = \lambda_{\text{final}} \cdot \mathcal{L}_{\text{final}} + \lambda_{\text{individual}} \cdot \mathcal{L}_{\text{individual}}
$$
where $\mathcal{L}_{\text{final}} = \text{MSE}(y, \hat{y})$ measures ensemble prediction accuracy, and $\mathcal{L}_{\text{individual}} = \sum_{i=1}^3 \text{MSE}(y, \hat{y}_i)$ encourages each branch to retain predictive utility. The loss weights $\lambda_{\text{final}}$ and $\lambda_{\text{individual}}$ control the balance between overall performance and individual modality contributions. Empirical evaluation confirms that joint fine-tuning of modality branches with the gating network consistently improves predictive accuracy over freezing branches after Phase 1.

Together, these design choices establish GAMENet as a robust and interpretable framework for multimodal popularity prediction, effectively combining domain-specific representations with adaptive and interpretable fusion network. 

\section{Experiments and Results}\label{sec:exp}

\subsection{Experimental Setup}
All experiments were conducted on the M4A-CTD dataset (53,777 tracks), using a fixed random seed of 46 for reproducibility. All model training was performed on a single NVIDIA A100 GPU (80GB RAM). For audio feature compression, we applied the OnionEnsembleAENet framework on the Music4All-ONION dataset, using the selected 11,851 features grouped into seven categories. Each group was independently standardized before training. The models were optimized using Adam with a base learning rate of $1 \times 10^{-4}$. Regularization included adaptive $L_2$ penalties, gradient clipping (norm 1.0), and early stopping. All models used a batch size of 256 with mixed-precision training. OnionEnsembleAENet reduced the audio feature space to 2,352 dimensions, with compression ratios varying from 11.4\% (BLF features) to 29.2\% (Small Combined group). The average Relative MSE (RelMSE) across feature groups was 0.175, with the Essentia group achieving the best reconstruction fidelity (RelMSE = 0.091).

During final feature data preprocessing, modality-specific normalization was applied. Compressed audio features were already standardized during autoencoder training. The 46 social metadata features were normalized using z-score scaling on the training set. Lyrics embeddings from OpenAI’s text-embedding-3-large model (3,072 dimensions) were scaled by $100 \times$ to match the scale of other modalities. The embeddings were obtained via OpenAI’s batch processing API, with a total cost of nearly \$5. Additionally, we initially selected a 65-dimensional sentiment feature group from the Music4All-ONION, however, it was dropped after empirical evaluation showed minimal downstream predictive performance. The final multimodal dataset contains 53,726 tracks with features across audio, social metadata, and lyrics. The combined feature vector has 5,470 dimensions. The popularity target was MinMax scaled to [0, 1]. Dataset splits followed the strategy described in \ref{subsec:data cleaning-preprocessing}. For experiments on the SPD\_cleaned dataset, we adopted the train-test split from prior work \cite{hitmusicnet}. Model performance was evaluated using standard regression metrics --- $R^2$, Mean Absolute Error (MAE), Mean Squared Error (MSE) --- along with Relative MSE (RelMSE) to assess variance capture on scaled popularity values.

\begin{table}[t]
\centering
\footnotesize
\setlength{\tabcolsep}{8pt}
\begin{tabular}{l@{\hspace{3pt}}ccc}
\toprule
\multirow{2}{*}{\textbf{Model}} &
\multicolumn{3}{c}{\textbf{Feature Set ($R^2$ scores)}} \\
\cmidrule(lr){2-4}
 & Acoustic & +CTD Agg. & +CTD Temp. \\ 
 & (8 feat.) & (18 feat.) & (45 feat.) \\ \midrule
LR            & 0.0685 & 0.3033 & 0.3634 \\
Random Forest & \textbf{0.1319} & 0.6027 & 0.6034 \\
LightGBM      & 0.0566 & 0.6984 & 0.7483 \\
XGBoost       & 0.0674 & \textbf{0.6999} & \textbf{0.7485} \\
Neural Net    & 0.0885 & 0.3576 & 0.4826 \\ 
\bottomrule
\end{tabular}
\caption{CTD feature ablation on Music4All popularity prediction using classical ML techniques and Neural Network.}
\label{tab:ctd_ablation}
\end{table}



\subsection{Ablation Study: Impact of CTD Features}\label{subsec: ctd ablation}

As introduced in \ref{subsec: Methodology CTD}, CTD features capture artist- and song-level listening patterns aggregated over five years (2016--2020), offering rich social signals.
Table~\ref{tab:ctd_ablation} summarizes the performance of standard ML regression models under three feature configurations: (i) Acoustic Descriptors (8 features) as a baseline; (ii) Acoustic + Aggregate CTD features (18 total features); and (iii) Acoustic + Aggregate + Temporal CTD features (45 total features). Including Aggregate CTD features leads to significant performance gains across all models. For instance, Random Forest improves from $R^2 = 0.132$ (acoustic only) to $R^2 = 0.603$ with aggregate CTD. Ensemble methods such as XGBoost and LightGBM reach $R^2$ values around 0.700. Adding Temporal CTD features yields further improvements. XGBoost attains $R^2 = 0.749$ when temporal dynamics are included --- a relative gain of $\sim$7\% over the aggregate-only scenario --- confirming the importance of modeling artist and song popularity trajectories over time.
These results validate our hypothesis that temporal listener dynamics, career progression, and momentum carry meaningful predictive signals, complementing static features in popularity prediction tasks.

\subsection{MultiModal Baseline on Music4All}
\begin{table}[t]
\centering
\small
\begin{tabular}{l l c c c r}
\toprule
\textbf{Feature Combination} & \textbf{Model} & \textbf{MAE} & $R^2$\\ \midrule
Base (CTD + Spotify)         & LightGBM & 0.0522 & 0.7483 \\
                             & XGBoost & 0.0525 & \textbf{0.7485} \\
                             & Neural Net & 0.0525 & 0.4826\\ \midrule
Base + Lyrics                & LightGBM & 0.0604 & \textbf{0.6953} \\
                             & XGBoost & 0.0626 & 0.6749          \\ \midrule
Base + Audio                 & LightGBM & 0.0662 & \textbf{0.6350} \\
                             & XGBoost & 0.0678 & 0.6207          \\ \midrule
All Modalities               & LightGBM& 0.0611 & \textbf{0.6879}\\
                             & XGBoost & 0.0638 & 0.6654         \\
                             & Neural Net & 0.0692 & 0.6012  \\ \bottomrule
\end{tabular}
\caption{Performance of ML models and deep neural networks on M4A-CTD with increasing feature complexity. We report MAE and $R^2$ on the test set for four feature combinations. }
\label{tab:music4all_baselines}
\end{table}

To establish a baseline for music popularity prediction on the Music4All family of datasets, we conducted experiments to assess both the dataset’s modeling potential and the comparative performance of traditional machine learning models versus deep neural networks. Given that, to the best of our knowledge, this is the first large-scale modeling of Music4All for popularity prediction. We evaluated four progressively expanded feature sets: (i) Base (CTD + Spotify Acoustic), consisting of 45 features; (ii) Base + Lyrics, adding the 3,072-dimensional OpenAI lyrics embeddings; (iii) Base + Audio, adding 2,352 compressed audio features obtained from OnionEnsembleAENet; and (iv) All Modalities --- 5,470-dimensional representation.

For traditional ML models, we used XGBoost and LightGBM, applying feature-aware hyperparameter tuning, including stronger regularization and lower learning rates for higher-dimensional input spaces. For deep learning, we designed a fully connected neural network with hidden layers [2048, 1024, 512, 128], ELU activations, batch normalization, and progressive dropout. The network was trained with Adam optimizer. The results, summarized in Table~\ref{tab:music4all_baselines}, reveal distinct trends. With the Base features, gradient boosting models performed strongly, achieving $R^2 \approx 0.75$, while the deep neural network started lower at $R^2 = 0.48$. However, as feature complexity increased, boosting models showed declining performance --- particularly with the addition of high-dimensional lyrics or audio features --- indicating their limited capacity to capture complex semantic relationships and multimodal interactions. In contrast, the deep neural network exhibited steady performance gains with each added modality, reaching $R^2 = 0.60$ with the full feature set. These findings motivated the design of an explicit deep learning architecture for multimodal learning and downstream popularity prediction.

\subsection{GAMENet Results}

\begin{table}[t]
\centering
\small
\setlength{\tabcolsep}{8pt}      
\renewcommand{\arraystretch}{1.05}  
\begin{tabular}{l l l c c}
\toprule
\textbf{Dataset} & \textbf{Phase} & \textbf{Modality} & \textbf{MAE} & $R^2$\\
\midrule
\multirow{5}{*}{M4A-CTD}
 & \multirow{3}{*}{I}
   & Audio             & 0.1132 & 0.2241\\
 & & Lyrics            & 0.1070 & 0.3073\\
 & & Social            & 0.0804 & \textbf{0.5714}\\
 \cmidrule(lr){2-5}
 & II
   & Ensemble & \textbf{0.0706} & \textbf{0.6761}\\
\addlinespace
\midrule
\multirow{5}{*}{SPD\_Cleaned}
 & \multirow{3}{*}{I}
   & Audio             & 0.1214 & 0.2543\\
 & & Lyrics            & 0.1022 & 0.4169\\
 & & Social            & 0.0820 & \textbf{0.6483}\\
 \cmidrule(lr){2-5}
 & II
   & Ensemble & \textbf{0.0735} & \textbf{0.7013}\\
\bottomrule
\end{tabular}
\caption{Phase-wise performance of \textsc{GAMENet} on M4A-CTD and the
SPD\_Cleaned dataset.  Phase I trains modality-specific experts; Phase II fine-tunes them jointly via adaptive gating.  Best $R^2$ within each dataset is
\textbf{bold}.}
\label{tab:gamenet_multidataset}
\end{table}

\textbf{Phase I: Modality-Expert Pretraining.} Individual modality experts for audio, lyrics, and social metadata were independently trained using Adam optimization, early stopping (patience=25), and a \texttt{ReduceLROnPlateau} learning-rate scheduler with batch size of 256 to predict music popularity score. Social metadata yielded the highest standalone performance ($R^2=0.571$), followed by lyrics ($R^2=0.307$) and audio features ($R^2=0.224$), indicating social signals' derived from historical listener engagement superior predictive capability.

\textbf{Phase II: Adaptive Gating Ensemble.} In the second phase, modality branches were fine-tuned jointly using a learnable gating mechanism designed to dynamically weight contributions from each modality based on intermediate representations. The gating network comprised a shallow multilayer perceptron with softmax-normalized outputs, trained with AdamW optimization at a learning rate of $5\times10^{-6}$ and batch size of 256. The gating network successfully learned to emphasize social metadata (average attention weight of 0.478), followed by lyrics (0.287), and audio features (0.235). Joint fine-tuning substantially improved predictive performance, yielding an ensemble $R^2$ of 0.676 (MAE=0.0706), an 18.4\% relative improvement over the best individual modality and a 12.6\% improvement over the baseline deep neural network ($R^2 = 0.60$) tested on the same multimodal input.

To further validate the robustness and generalizability of GAMENet, we trained and evaluated our model on the larger and independent SPD\_Cleaned dataset (74,206 tracks). On this dataset, GAMENet achieved an ensemble $R^2$ score of 0.701 (MAE=0.0735). When compared against HitmusicNet\footnote{Code: \url{https://github.com/dmgutierrez/hitmusicnet}}, a state-of-the-art multimodal popularity prediction baseline previously evaluated on SPD\_Cleaned, GAMENet achieved a 16\% improvement in MAE (0.0735 vs.\ 0.0877) demonstrating robustness, and predictive generalizability.

\textbf{Gating Patterns and Error Analysis} The learned gate values in GAMENet reliably captured the relative importance of each modality. On the training set, social metadata received the highest average weight (0.483), followed by lyrics (0.284) and audio (0.233). These proportions remained consistent on the test set (0.478, 0.287, 0.235), indicating stable generalization without overfitting. A decade-wise breakdown showed that social metadata consistently dominated, though its weight declined slightly after 2000, with modest gains for lyrics and audio—suggesting a growing relevance of content features in recent music. Prediction error analysis confirmed model calibration: predicted popularity distributions matched the actual ones in both mean and spread, and residuals showed no skew or bias. This indicates that GAMENet not only achieved strong predictive accuracy but also captured the variance structure of popularity across time.

\section{Conclusion}\label{sec: conclusion}

In this work, we presented GAMENet, a multimodal deep learning framework for predicting music popularity by adaptively combining content features and listener engagement signals. To support this, we introduced M4A-CTD, a curated subset of the Music4All corpus enriched with temporally structured listening logs and systematically engineered Career Trajectory Dynamics (CTD) features. Our design captures both short-term trends at the song level and long-term career dynamics at the artist level, addressing key gaps in prior work related to static modeling and oversimplified fusion strategies. Extensive experiments on Music4All and SpotGenTrack show that GAMENet outperforms existing baselines and that CTD features contribute substantial predictive value. These results underscore the importance of temporal structure and social feedback in modeling cultural success. Looking ahead, our framework opens promising directions for integrating real-time engagement signals, modeling cross-platform influence, and studying the temporal evolution of music trends at scale.
\bibliography{aaai25}

@article{Yee22,
author = {Yee, Yap and Raheem, Mafas},
year = {2022},
month = {09},
pages = {1786-1799},
title = {Predicting Music Popularity Using Spotify and YouTube Features},
volume = {15},
journal = {Indian Journal Of Science And Technology},
doi = {10.17485/IJST/v15i36.2332}
}

@article{Li2021LSTMRPAAS,
  title={LSTM-RPA: A Simple but Effective Long Sequence Prediction Algorithm for Music Popularity Prediction},
  author={Kun Li and Meng-Jie Li and Yanling Li and Min Lin},
  journal={ArXiv},
  year={2021},
  volume={abs/2110.15790},
  url={https://api.semanticscholar.org/CorpusID:240288440}
}

@inproceedings{Araujo2017,
author = {Soares Araujo, Carlos Vicente and Mendonca Neto, Rayol and Nakamura, Fabiola and Nakamura, Eduardo},
year = {2017},
month = {10},
pages = {149-156},
title = {Predicting Music Success Based on Users' Comments on Online Social Networks},
doi = {10.1145/3126858.3126885}
}

@inproceedings{Lee2015,
author = {Lee, Junghyuk and Lee, Jong-Seok},
title = {Predicting Music Popularity Patterns based on Musical Complexity and Early Stage Popularity},
year = {2015},
isbn = {9781450337496},
publisher = {Association for Computing Machinery},
address = {New York, NY, USA},
url = {https://doi.org/10.1145/2802558.2814645},
doi = {10.1145/2802558.2814645},
booktitle = {Proceedings of the Third Edition Workshop on Speech, Language \& Audio in Multimedia},
pages = {3–6},
numpages = {4},
location = {Brisbane, Australia},
series = {SLAM '15}
}

@article{Chon2006,
author = {Chon, Song Hui and Slaney, Malcolm and Berger, Jonathan},
year = {2006},
month = {10},
pages = {},
title = {Predicting success from music sales data: a statistical and adaptive approach},
doi = {10.1145/1178723.1178736}
}

@article{Reisz2024Quantifying,
	author = {Reisz, Niklas and Servedio, Vito D. P. and Thurner, Stefan},
	journal = {Scientific Reports},
	doi = {10.1038/s41598-024-58969-w},
	issn = {2045-2322},
	number = {1},
	year = {2024},
	month = {apr 18},
	publisher = {{Springer Science and Business Media LLC}},
	title = {Quantifying the impact of homophily and influencer networks on song popularity prediction},
	url = {http://dx.doi.org/10.1038/s41598-024-58969-w},
	volume = {14},
}

@article{Zangerle2019Hit,
	author = {Zangerle, Eva and V{\" o}tter, M. and Huber, Ramona and Yang, Yi-Hsuan},
	journal = {International Society for Music Information Retrieval Conference},
	year = {2019},
	title = {Hit {Song} {Prediction}: Leveraging {Low}- and {High}-{Level} {Audio} {Features}},
}

@inproceedings{Silva2021Collaboration,
	author = {Silva, Mariana O. and Moro, Mirella M.},
	booktitle = {Anais {Estendidos} do {XXVII} {Simp}{\' o}sio {Brasileiro} de {Sistemas} {Multim}{\' i}dia e {Web} ({WebMedia} 2021)},
	doi = {10.5753/webmedia_estendido.2021.17603},
	year = {2021},
	month = {nov 5},
	pages = {11--14},
	organization = {Sociedade Brasileira de Computa{\c c}{\~ a}o - SBC},
	title = {Collaboration-{Aware} {Hit} {Song} {Analysis} and {Prediction}},
	url = {http://dx.doi.org/10.5753/webmedia_estendido.2021.17603},
}

@article{Vavaroutsos2024HSP,
	author = {Vavaroutsos, Petros and Vikatos, Pantelis},
	journal = {Science Talks},
	doi = {10.1016/j.sctalk.2024.100363},
	issn = {2772-5693},
	year = {2024},
	month = {6},
	pages = {100363},
	publisher = {Elsevier BV},
	title = {HSP-{TL}: A deep metric learning model with triplet loss for hit song prediction using lyrics and audio features},
	url = {http://dx.doi.org/10.1016/j.sctalk.2024.100363},
	volume = {10},
}

@article{Matsumoto2020Context,
	author = {Matsumoto, Yui and Harakawa, Ryosuke and Ogawa, Takahiro and Haseyama, Miki},
	journal = {IEEE Access},
	doi = {10.1109/access.2020.2978281},
	issn = {2169-3536},
	year = {2020},
	pages = {48673--48685},
	publisher = {{Institute of Electrical and Electronics Engineers (IEEE)}},
	title = {Context-{Aware} {Network} {Analysis} of {Music} {Streaming} {Services} for {Popularity} {Estimation} of {Artists}},
	url = {http://dx.doi.org/10.1109/ACCESS.2020.2978281},
	volume = {8},
}

@article{Yu2017Hit,
	author = {Yu, Lang-Chi and Yang, Yi-Hsuan and Hung, Yun-Ning and Chen, Yian},
	journal = {arXiv.org},
	year = {2017},
	title = {Hit {Song} {Prediction} for {Pop} {Music} by {Siamese} {CNN} with {Ranking} {Loss}},
}

@article{Vötter2022,
author = {Vötter, Michael and Mayerl, Maximilian and Specht, Guenther and Zangerle, Eva},
year = {2022},
month = {05},
pages = {1-23},
title = {HSP Datasets: Insights on Song Popularity Prediction},
volume = {16},
journal = {International Journal of Semantic Computing},
doi = {10.1142/S1793351X22400104}
}

@book{Celma2010,
author = {Celma, Oscar},
year = {2010},
month = {01},
pages = {},
title = {Music Recommendation and Discovery: The Long Tail, Long Fail, and Long Play in the Digital Music Space},
isbn = {978-3-642-13286-5},
journal = {Music Recommendation and Discovery: The Long Tail, Long Fail, and Long Play in the Digital Music Space},
doi = {10.1007/978-3-642-13287-2}
}

@INPROCEEDINGS{Bertin2011,
  author = {Thierry Bertin-Mahieux and Daniel P.W. Ellis and Brian Whitman and Paul Lamere},
  title = {The Million Song Dataset},
  booktitle = {{Proceedings of the 12th International Conference on Music Information
	Retrieval ({ISMIR} 2011)}},
  year = {2011}
}

@article{Shulman2016,
author = {Shulman, Benjamin and Sharma, Amit and Cosley, Dan},
year = {2016},
month = {03},
pages = {},
title = {Predictability of Popularity: Gaps between Prediction and Understanding},
volume = {10},
journal = {Proceedings of the International AAAI Conference on Web and Social Media},
doi = {10.1609/icwsm.v10i1.14748}
}

@inproceedings{Araujo2019,
author = {Araujo, Carlos and Cristo, Marco and Giusti, Rafael},
year = {2019},
month = {12},
pages = {859-864},
title = {Predicting Music Popularity Using Music Charts},
doi = {10.1109/ICMLA.2019.00149}
}

@inproceedings{Pachet2008,
author = {Pachet, Francois and Roy, Pierre},
year = {2008},
month = {01},
pages = {355-360},
title = {Hit Song Science Is Not Yet a Science.}
}

@inproceedings{Dhanaraj2005,
  title={Automatic Prediction of Hit Songs},
  author={Ruth Dhanaraj and Beth Logan},
  booktitle={International Society for Music Information Retrieval Conference},
  year={2005},
  url={https://api.semanticscholar.org/CorpusID:13024981}
}

@article{Seufitelli2023HitSS,
  title={Hit song science: a comprehensive survey and research directions},
  author={Danilo B. Seufitelli and Gabriel P. Oliveira and Mariana O. Silva and Clarisse Scofield and Mirella M. Moro},
  journal={Journal of New Music Research},
  year={2023},
  volume={52},
  pages={41 - 72},
  url={https://api.semanticscholar.org/CorpusID:265315926}
}

@ARTICLE{hitmusicnet,
  author={Martín-Gutiérrez, David and Hernández Peñaloza, Gustavo and Belmonte-Hernández, Alberto and Álvarez García, Federico},
  journal={IEEE Access}, 
  title={A Multimodal End-to-End Deep Learning Architecture for Music Popularity Prediction}, 
  year={2020},
  volume={8},
  number={},
  pages={39361-39374},
  doi={10.1109/ACCESS.2020.2976033}}

@inproceedings{Music4All-Onion,
author = {Moscati, Marta and Parada-Cabaleiro, Emilia and Deldjoo, Yashar and Zangerle, Eva and Schedl, Markus},
title = {Music4All-Onion -- A Large-Scale Multi-faceted Content-Centric Music Recommendation Dataset},
year = {2022},
isbn = {9781450392365},
publisher = {Association for Computing Machinery},
address = {New York, NY, USA},
url = {https://doi.org/10.1145/3511808.3557656},
doi = {10.1145/3511808.3557656},
booktitle = {Proceedings of the 31st ACM International Conference on Information \& Knowledge Management},
pages = {4339–4343},
numpages = {5},
location = {Atlanta, GA, USA},
series = {CIKM '22}
}

@article{Santana2020Music4AllAN,
  title={Music4All: A New Music Database and Its Applications},
  author={Igor Andr{\'e} Pegoraro Santana and Fabio Pinhelli and Juliano Donini and Leonardo Gabiato Catharin and Rafael B. Mangolin and Yandre M. G. Costa and Val{\'e}ria Delisandra Feltrim and Marcos Aur{\'e}lio Domingues},
  journal={2020 International Conference on Systems, Signals and Image Processing (IWSSIP)},
  year={2020},
  pages={399-404},
  url={https://api.semanticscholar.org/CorpusID:220734872}
}

\end{document}